\documentclass[lettersize,letterpaper]{IEEEtran}
\usepackage{amsmath,amsfonts}
\usepackage{algorithmic}
\usepackage{algorithm}
\usepackage{array}
\usepackage{textcomp}
\usepackage{stfloats}
\usepackage{url}
\usepackage{verbatim}
\usepackage{graphicx}
\usepackage{cite}
\usepackage{bm}
\usepackage{subcaption}
\usepackage{booktabs} 
\usepackage{multirow}
\usepackage{amsthm} 
\usepackage{hyperref} 
\usepackage{cleveref} 
\usepackage{makecell}

\begin{document}
	
	\title{RL based Beamforming Optimization for 3D Pinching Antenna assisted ISAC Systems}
	
	\author{Qian Gao,~\IEEEmembership{Graduate Student Member,~IEEE,} 
		Ruikang Zhong,~\IEEEmembership{Member,~IEEE,} \\
		Yue Liu,~\IEEEmembership{Member,~IEEE,}
		Hyundong Shin,~\IEEEmembership{Fellow,~IEEE,}
		Yuanwei Liu,~\IEEEmembership{Fellow,~IEEE}

		\thanks{Qian Gao, and Ruikang Zhong are with the School of Electronic Engineering and Computer Science, Queen Mary University of London, London, E1 4NS, U.K. (e-mail: \{q.gao, r.zhong\}@qmul.ac.uk).
		
		Yue Liu is with the Faculty of Applied Sciences, Macao Polytechnic
		University, Macao SAR, China. (e-mail: yue.liu@mpu.edu.mo)
		
		Hyundong Shin is with the Department of Electronics and Information
		Convergence Engineering, Kyung Hee University, 1732 Deogyeong-daero,
		Giheung-gu, Yongin-si, Gyeonggi-do 17104, Republic of Korea. (e-mail:
		hshin@khu.ac.kr)

		Yuanwei Liu  is with
		the Department of Electrical and Electronic Engineering, The University
		of Hong Kong, Hong Kong. (e-mail: yuanwei@hku.hk).
	
}
	}

	\maketitle

\begin{abstract}
In this paper, a three-dimensional (3D) deployment scheme of pinching antenna array is proposed, aiming to enhances the performance of integrated sensing and communication~(ISAC) systems. To fully realize the potential of 3D deployment, a joint antenna positioning, time allocation and transmit power optimization problem is formulated to maximize the sum communication rate with the constraints of target sensing rates and system energy. To solve the sum rate maximization problem, we propose a heterogeneous graph neural network based reinforcement learning (HGRL) algorithm. Simulation results prove that 3D deployment of pinching antenna array outperforms 1D and 2D counterparts in ISAC systems. Moreover, the proposed HGRL algorithm surpasses other baselines in both performance and convergence speed due to the advanced observation construction of the environment.

\end{abstract}

\begin{IEEEkeywords}
Beamforming, graph neural network (GNN), integrated sensing and communication (ISAC), reinforcement learning (RL), pinching antenna.
\end{IEEEkeywords}

\section{Introduction}

Integrated sensing and communication (ISAC) has emerged as a transformative paradigm in next-generation wireless systems \cite{ISAC6G}. By enabling spectrum and hardware sharing between these two traditionally separate functions, ISAC systems meet requirements of increased spectral efficiency, reduced infrastructure cost, and enhanced situational awareness. Such capabilities are crucial for emerging applications such as autonomous vehicles \cite{V2X,UAVXX}, and WiFi sensing~\cite{hs}.

Antennas play a vital role in the resource utilization efficiency of ISAC. Various flexible antennas have been proposed to promote the adaptability and performance of ISAC systems. However, existing flexible antennas \cite{move, ris, fuild} are with limited antenna variation and their utilization cost is high when antenna number is large. Fortunately, pinching antennas \cite{PA} having the low-cost and highly reconfigurable features became a potential candidate antenna technology for the ISAC system.  As dielectric-waveguide-based leaky antennas, pinching antennas can be activated at any position by “pinching” the waveguide, forming on-demand line-of-the-sight (LoS) beams to both users and sensing targets without expensive phase-array hardware.  This mobility counters large-scale path loss, enlarges the effective aperture, and naturally supports near-field spherical-wave operation, all of which are beneficial to ISAC throughput and sensing accuracy.

Existing ISAC studies \cite{PA1Ddl, PA1Dul, PA2D} for pinching antennas place all elements on a single straight waveguide or on a pair of parallel waveguides lying in the \(x\!-\!y\) plane.  Although these 1D/2D deployments simplify analysis, they lose elevation diversity, and limit angular coverage. In this paper, we extend existing pinching antenna assisted ISAC system to the 3D configuration.  
Compared with 1D/2D layouts, the proposed 3D deployment not only offers full-space beam steering for airborne or elevated users and targets but also provides additional spatial degrees-of-freedom (DoFs) that decouple communication and sensing beams. 
However, the 3D geometry of pinching antennas introduces a larger continuous action space (antenna displacement, TDMA weights, and power allocation) that conventional reinforcement learning (RL) or hand-crafted optimization can barely handle \cite{graphpinch}.  
We therefore propose HGRL, a heterogeneous graph neural network based reinforcement learning framework that  
firstly represents the antenna–user–target topology as a time-varying heterogeneous graph and encodes it via a graph neural network (GNN) \cite{GNN};  
then utilises a RL policy to jointly adjust 3D antenna positioning, TDMA allocation and transmit power assignment.

The main contributions of this paper can be summarized as below:
\begin{itemize}
	\item We formulate the 3D pinching antenna assisted ISAC system as a Markov decision process (MDP) that captures fine-grained resource control. The objective is to maximize the sum communication rate, subject to the target sensing rate and transmit power budget.
	\item We develop HGRL, the first heterogeneous GNN based RL framework for 3D pinching antenna assisted ISAC system, enabling scalable optimization over graph-structured states and mixed action spaces.
	\item We conduct numerical experiments to demonstrate the
	effectiveness of the proposed 3D pinching antennas and HGRL algorithm in enhancing the ISAC performance. 
\end{itemize}

The rest of this paper is organized as follows: Section~\ref{section:2} illustrates the system model, and problem formulation. In Section \ref{section:3}, we propose the HGRL algorithm for antenna positioning and transmit power optimization. Then, the numerical results and performance analysis are provided in Section \ref{section:4}. Finally, we conclude the paper in Section \ref{section:5}.

\section{System Model and Problem Formulation} \label{section:2}

\begin{figure}[t!]
	\centering
	\captionsetup{justification=centering}
	\includegraphics[width=0.4\textwidth]{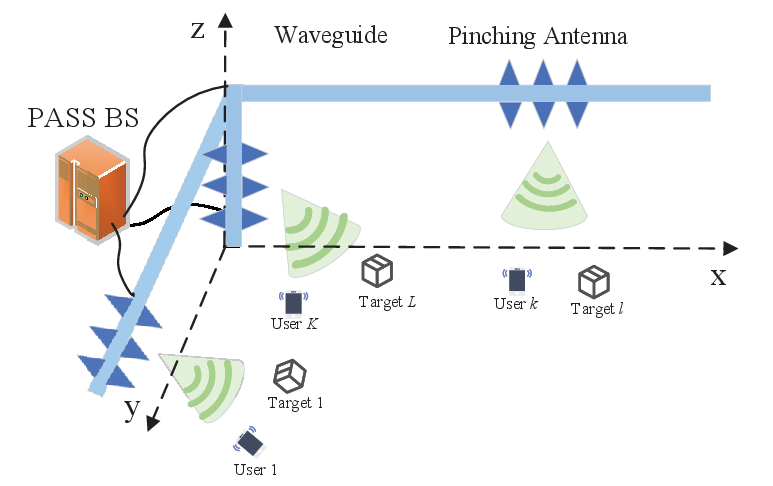}
	\caption{Illustration of the 3D pinching antenna assisted downlink ISAC system.}
	\label{Fig.1}
\end{figure}

In this paper, we propose a 3D pinching antenna assisted downlink ISAC system, as shown in Fig. \ref{Fig.1}. We assume that base station~(BS) is equipped with three orthogonal waveguides, each aligned along one of the Cartesian axes ($x$, $y$, $z$), to jointly serve $K$ single-antenna communication users and detect $L$ sensing targets. Each waveguide contains $N$ pinching antennas~(PAs), resulting in $M = 3N$ total antennas. The positions of the $n$th PA on the $x$-, $y$-, and $z$-aligned waveguides are denoted as $\psi^{p}_{n_x} = (x_n, 0, d)$, $\psi^{p}_{n_y} = (0, y_n, d)$, and $\psi^{p}_{n_z} = (0, 0, z_n)$, respectively, where $d$ is the fixed elevation above the $x$-$y$ plane. Communication users and sensing targets are located in the $x$-$y$ plane, with coordinates $\psi^{\text{com}}_k = (x_k, y_k, 0)$ and $\psi^{\text{sen}}_l = (x_l, y_l, 0)$, respectively.

Due to the limited number of waveguides, we adapt a time-division multiple access (TDMA) scheme to separately serve communication users and sensing targets. Specifically, each time slot $t \in \mathcal{T} = \{1,2,\dots,T\}$ is first divided into different communication proportions $q^{\text{com}}_{k,t}$, and they satisfy $\sum_{k=1}^{K}q^{\text{com}}_{k,t}  \leq 1$. Then, targets can be sensed during these proportions by considering communication symbol as radar pulse. Moreover, the configuration of the system is assumed to be fixed during time slot $t$, but it varies across different slots to optimize the ISAC performance.

\subsection{Signal Model}

The transmitted signal at the BS in time slot $t$ is composed of a combination of the communication signal $s^{\text{com}}_{t,k}$
\begin{align}
	s_t = \sum_{k=1}^{K}z_{k} s^{\text{com}}_{k,t} ,
\end{align}
where the signal energy is normalized $\quad \mathbb{E}[|s_t|^2] = 1$ and $[z_1,  \cdots, z_k,\cdots, z_K]$ is a one-hot vector.

Then, this signal is fed into waveguides at the same time for signal radiation. For each waveguide, the signals transmitted by PAs are phase-shifted version of each other. The radiation signal can be expressed as

\begin{align}
	\mathbf{s}_t = \left[ \sqrt{\frac{P}{M}}e^{-j\theta_{1,t}}, \cdots,  \sqrt{\frac{P}{M}}e^{-j\theta_{N,t}} \right]^{T} s_t,
\end{align}
where $P$ is the transmit power, $\theta_{n,t}$ is the signal phase shift at $n$-th PA and $\theta_{n,t} = 2\pi \frac{|\psi^{p}_{0}-\psi^{p}_{n_a,t}|}{\lambda_0}$. Thereof, $\psi^{p}_{0}$is the location of feed point, $a \in \{x, y, z\}$ denotes the waveguide parallel to different axis, $\lambda_0 = \frac{\lambda}{n_{\text{neff}}}$ is the waveguide wavelength in a dielectric waveguide, $n_{\text{neff}}$ is the effective refractive index and $\lambda$ represents wavelength.

Since the employment of PAs provides LoS channel condition for ISAC, the near-field spherical wave channels for communication user $k$ and sensing target $l$ can be expressed as 
\begin{align}
	\mathbf{h}_{k,t} = \left[\frac{\alpha e^{-j \frac{2\pi}{\lambda}|\psi^{\text{com}}_{k,t} -\psi^{p}_{1_a, t}|}}{|\psi^{\text{com}}_{k,t}  -\psi^{p}_{1_a, t}|}, \cdots,
	 \frac{\alpha e^{-j \frac{2\pi}{\lambda}|\psi^{\text{com}}_{k,t}  -\psi^{p}_{N_a, t}|}}{|\psi^{\text{com}}_{k,t} -\psi^{p}_{N_a, t}|}\right],
\end{align}
and
\begin{align}
	\mathbf{h}_{l,t} =  \left[\frac{\alpha e^{-j \frac{2\pi}{\lambda}|\psi^{\text{sen}}_{l,t} -\psi^{p}_{1_a, t}|}}{|\psi^{\text{sen}}_{l,t}  -\psi^{p}_{1_a, t}|}, \cdots,
	\frac{\alpha e^{-j \frac{2\pi}{\lambda}|\psi^{\text{sen}}_{l,t}  -\psi^{p}_{N_a, t}|}}{|\psi^{\text{sen}}_{l,t} -\psi^{p}_{N_a, t}|}\right],
\end{align}
where $\alpha = \frac{c}{4\pi f_c}$ is a constant that depends on the speed of light $c$ and the carrier frequency $f_c$, $\psi^{\text{com}}_{k,t}$ and $\psi^{\text{sen}}_{l,t}$ denote the location of user $k$ and target $l$ in time slot $t$, respectively.

\subsection{Communication and Sensing Metrics}

The received signal at communication user $k$ in slot $t$ is $y_{k,t} = \mathbf{h}_{k,t}\mathbf{s}_t + n_{k,t}$, which can be rewritten as 
\begin{align}
	y_{k,t} =\left( \sum_{n=1}^{M}\frac{\alpha e^{-j \frac{2\pi}{\lambda}|\psi^{\text{com}}_{k,t} -\psi^{p}_{n_a, t}|}}{|\psi^{\text{com}}_{k,t}  -\psi^{p}_{n_a, t}|} e^{-j\theta_{n,t}} \right) \sqrt{\frac{p_{k,t}}{M}} s_t + n_{k,t},
\end{align}
where $p_{k,t}$ is the transmit power for user $k$ and $n_{k,t}$ represents the additive Gaussian noise.

Then, the data rate of user $k$ can be expressed as
\begin{align}
	R^{\text{com}}_{k,t} &= q^{com}_{k,t} \log_2 \left(1+\frac{p_{k,t}}{Mq^{com}_{k,t} \sigma^2} \right. \notag \\
	&\left. \times  \left| \sum_{n=1}^{M}\frac{\alpha e^{-j \frac{2\pi}{\lambda}|\psi^{\text{com}}_{k,t} -\psi^{p}_{n_a, t}|}}{|\psi^{\text{com}}_{k,t}  -\psi^{p}_{n_a, t}|} e^{-j\theta_{n,t}}  \right|^2\right),
\end{align}
where $\sigma^2$ is the noise power. Following \cite{follow}, we use the signal-to-noise (SNR) received by targets to substitute the SNR of echo signal received by BS in this paper. The SNR of target $l$ in time slot $t$ can be denoted by 
\begin{align}
	\Gamma_{l,t} =\sum_{k=1}^{K} \frac{\left| \sum_{n=1}^{M}\frac{\alpha e^{-j \frac{2\pi}{\lambda}|\psi^{\text{sen}}_{l,t} -\psi^{p}_{n_a, t}|}}{|\psi^{\text{sen}}_{l,t}  -\psi^{p}_{n_a, t}|} \frac{p_{k,t}}{M} e^{-j\theta_{n,t}}  \right|^2}{\left(\left| \sum_{n=1}^{M}\frac{\alpha e^{-j \frac{2\pi}{\lambda}|\psi^{\text{com}}_{k,t} -\psi^{p}_{n_a, t}|}}{|\psi^{\text{com}}_{k,t}  -\psi^{p}_{n_a, t}|} \frac{p_{k,t}}{M} e^{-j\theta_{n,t}}  \right|^2 + \sigma^2\right)}.
\end{align}

\subsection{Problem Formulation}

Based on the above system model, we formulate the joint optimization of the pinching antenna positions $\{\psi^p_{n_a,t}\}$, the TDMA time allocations $\{q^{\text{com}}_{k,t}\}$, and the transmit powers $\{p_{k,t}\}$ to maximize the total communication data rate across all users and time slots.
The optimization problem is:
\begin{subequations} \label{eq:opt_problem}
	\begin{align}
		\max_{\{\psi^p_{n_a,t}, q^{\text{com}}_{k,t}, p_{k,t}\}} &\quad \sum_{t=1}^{T} \sum_{k=1}^{K} R^{\text{com}}_{k,t} \notag \\
		\text{s.t.}\quad 
		&\quad \Gamma_{l,t} \geq \Gamma_{\min}, \quad \forall l \in \{1,...,L\},\forall t, \label{eq:opt_problem_a} \\
		&\quad \sum_{k=1}^{K} q^{\text{com}}_{k,t} \leq 1, \quad \forall t, \label{eq:opt_problem_b} \\
		&\quad \sum_{t=1}^{T} \sum_{k=1}^{K} p_{k,t} q^{\text{com}}_{k,t} \leq E, \label{eq:opt_problem_c} \\
		&\quad 0 \leq p_{k,t} \leq p_{\max}, \quad \forall k,t, \label{eq:opt_problem_d} \\
		&\quad \|\psi^{p}_{i_a,t} - \psi^{p}_{j_a,t}\| \geq \delta, \quad \forall i \neq j, \forall a,t. \label{eq:opt_problem_e}
	\end{align}
\end{subequations}
where $p_{\max}$ denotes the maximum transmit power for each user. Constraint (8a) ensures the minimum SNR requirement for target sensing, constraints (8b) and (8c) enforce the total transmit power and overall energy budget limits, respectively. Constraint (8d) guarantees the minimum spacing between any two pinching antennas. This optimization problem is nonlinear and non-convex due to the presence of fractional SNR expressions, absolute values, and exponential phase terms.

\section{Heterogeneous Graph Neural Network based Reinforcement Learning Solution }\label{section:3}

In this section, we introduce HGRL, a novel reinforcement learning framework designed to jointly optimize pinching antenna positioning, TDMA time allocation, and transmit power control in 3D pinching antenna-assisted downlink ISAC systems. Unlike conventional homogeneous GNN-based methods, HGRL leverages a heterogeneous graph structure to explicitly encode distinct node types (antennas, users, and sensing targets) and relation types (communication, sensing, and interference). This heterogeneous design enhances the representational power of the agent by allowing more precise modeling of ISAC-specific interactions.

\subsection{MDP Formulation}

We formulate the optimization problem as a standard Markov Decision Process (MDP), where the agent interacts with the environment at every time step.

\textbf{State Space $\mathcal{S}$:} The state includes the heterogeneous graph $\mathcal{G}_t = (\mathcal{V}, \mathcal{E})$ representing the system topology at time $t$, where nodes $\mathcal{V}$ include antennas, users, and sensing targets, and edges $\mathcal{E}$ represent relation types. Node features include type indicators, 3D positions, and contextual communication/sensing relevance. The graph is encoded into node and global embeddings using a heterogeneous GNN encoder $\phi_{\text{HetGNN}}$.

\textbf{Action Space $\mathcal{A}$:} The agent outputs a continuous control action $a_t = [\Delta\bm{\psi}^p_{t}, \bm{q}^{\text{com}}_{t}, \bm{p}_{t}]$, representing antenna displacement, TDMA allocation, and power control.

\textbf{Reward Function $\mathcal{R}_t$:} The reward encourages joint ISAC performance while penalizing violations:
\begin{equation}
	\mathcal{R}_t = \sum_{k=1}^K R^{\text{com}}_{k,t} - \lambda_1 \sum_{l=1}^L \max(0, \Gamma_{\min} - \Gamma_{l,t}) - \lambda_2 C_{\text{phys}}
\end{equation}
where $\Gamma_{l,t}$ is the sensing SNR for target $l$ and $C_{\text{phys}}$ penalizes physical violations such as antenna spacing constraints.

\subsection{Heterogeneous Graph-Based Representation Learning}

At each time step $t$, the system is represented as a heterogeneous undirected graph $\mathcal{G}_t = (\mathcal{V}, \mathcal{E})$ with node types $\mathcal{T}_v = \{\text{antenna}, \text{user}, \text{target}\}$ and edge types $\mathcal{T}_e = \{\text{communicates}, \text{senses}, \text{interference}\}$. The node feature for each $v \in \mathcal{V}$ is defined as
\begin{equation}
	\mathbf{x}_v = [\tau_v, \mathbf{p}_v] \in \mathbb{R}^d
\end{equation}
where $\tau_v$ is a one-hot encoding of the node type, $\mathbf{p}_v \in \mathbb{R}^3$ is the 3D position.
Using a relation-aware GNN, the node embedding update is performed as
\begin{equation}
	\mathbf{h}^{(l+1)}_v = \sigma\left(\sum_{r \in \mathcal{T}_e} \sum_{u \in \mathcal{N}_r(v)} W_r^{(l)} \mathbf{h}^{(l)}_u + W_0^{(l)} \mathbf{h}^{(l)}_v\right)
\end{equation}
where $\mathcal{N}_r(v)$ is the set of neighbors connected to $v$ via relation $r$, $W_r^{(l)}$ are learnable weight matrices, and $\sigma$ is an activation function.

A global graph representation is obtained by mean-pooling:
\begin{equation}
	\mathbf{h}_t = \frac{1}{|\mathcal{V}|} \sum_{v \in \mathcal{V}} \mathbf{h}_{v,t}
\end{equation}
This $\mathbf{h}_t$ is then fed into the policy and value networks.

\begin{algorithm}[ht]
	\caption{HGRL: Heterogeneous Graph RL for ISAC Optimization}
	\label{alg1}
	\begin{algorithmic}[1]
		\STATE Initialize heterogeneous GNN encoder $\phi_{\text{HetGNN}}$, policy $\pi_\theta$, critic $V_\psi$.
		\FOR{each episode}
		\STATE Initialize environment and construct heterogeneous graph $\mathcal{G}_0$.
		\FOR{each step $t$}
		\STATE Encode graph: $\mathbf{h}_t \leftarrow \phi_{\text{HetGNN}}(\mathcal{G}_t)$.
		\STATE Sample action: $a_t \sim \pi_\theta(a_t | \mathbf{h}_t)$.
		\STATE Apply action $a_t$, observe reward $r_t$ and $\mathcal{G}_{t+1}$.
		\STATE Store transitions.
		\STATE Update $V_\psi$, $\pi_\theta$ using A2C objectives.
		\ENDFOR
		\ENDFOR
	\end{algorithmic}
\end{algorithm}

\subsection{HGRL Algorithm}

In our proposed method, we adopt the Advantage Actor-Critic (A2C) algorithm  \cite{A2C} to optimize the agent's decision-making policy. A2C is a synchronous, on-policy reinforcement learning algorithm that strikes a balance between sample efficiency and training stability. It utilizes both a policy network (actor) and a value network (critic), which are updated jointly to reduce variance in policy gradient estimates.

To effectively capture the complex topological and relational information in our environment, we employ a Graph Convolutional Network (GCN) \cite{GCN} to extract structural features from the heterogeneous graph. These graph-based embedded outputs are then used as input to the actor-critic framework, enabling more informed and spatially-aware decision making. In this algorithm, the agent consists of:

\paragraph{Policy Network $\pi_\theta$}
A continuous actor network that outputs actions given heterogeneous GNN embeddings: $a_t \sim \pi_\theta(a_t|\mathbf{h}_t)$. It is optimized using A2C with the clipped objective:
\begin{equation}
	\small
	\mathcal{L}^{\text{CLIP}}(\theta) = \mathbb{E}_t \left[ \min \left( r_t(\theta) \hat{A}_t, \ \text{clip}(r_t(\theta), 1 - \epsilon, 1 + \epsilon) \hat{A}_t \right) \right],
\end{equation}
where $r_t(\theta) = \frac{\pi_\theta(a_t|s_t)}{\pi_{\text{old}}(a_t|s_t)}$ and $\hat{A}_t$ is the advantage estimate.

\paragraph{Value Network $V_\psi$}
The critic network estimates expected return and is trained by minimizing the TD error:
\begin{equation}
	\mathcal{L}_V = \mathbb{E}_t \left[ \left( V(s_t) - \hat{R}_t \right)^2 \right].
\end{equation}

\paragraph{Graph Encoder $\phi_{\text{HetGNN}}$}
A heterogeneous GNN module that computes structured representations from the input graph.

\section{Simulation Results}\label{section:4}

In this section, we conduct experimental simulations for the proposed 3D pinching antenna deployment and HGRL algorithm. In the simulation, we consider a multi-user ISAC system with a configurable pinching antenna deployment geometry, where $U = 6$ users and $T = 1$ target are randomly distributed within a $50\,\mathrm{m} \times 50\,\mathrm{m}$ area at ground level. The antenna array consists of $N = 6$ elements. The carrier frequency is set to $f = 28\,\mathrm{GHz}$, and each antenna element is allowed to adjust its position under a minimum inter-element spacing $\delta =\lambda/2$. Communication channels are modeled as LoS paths with effective refractive index $n_{\mathrm{eff}} = 1.4$. The TDMA time fractions and power allocations are dynamically optimized for each user, subject to a total power constraint $P_{\max} = 100\,\mathrm{W}$ and a maximum per-antenna power limit of $0.1\,\mathrm{W}$. The communication data rate is computed based on coherent beamforming with additive white Gaussian noise power of $-90\,\mathrm{dBm}$, and the sensing performance is evaluated via the maximum SNR achieved for the target. The sensing threshold is set to $10\,\mathrm{dB}$.

\begin{figure}[htbp]
	\centering
	\begin{subfigure}[t]{0.22\textwidth}
		\centering
		\includegraphics[width=\linewidth]{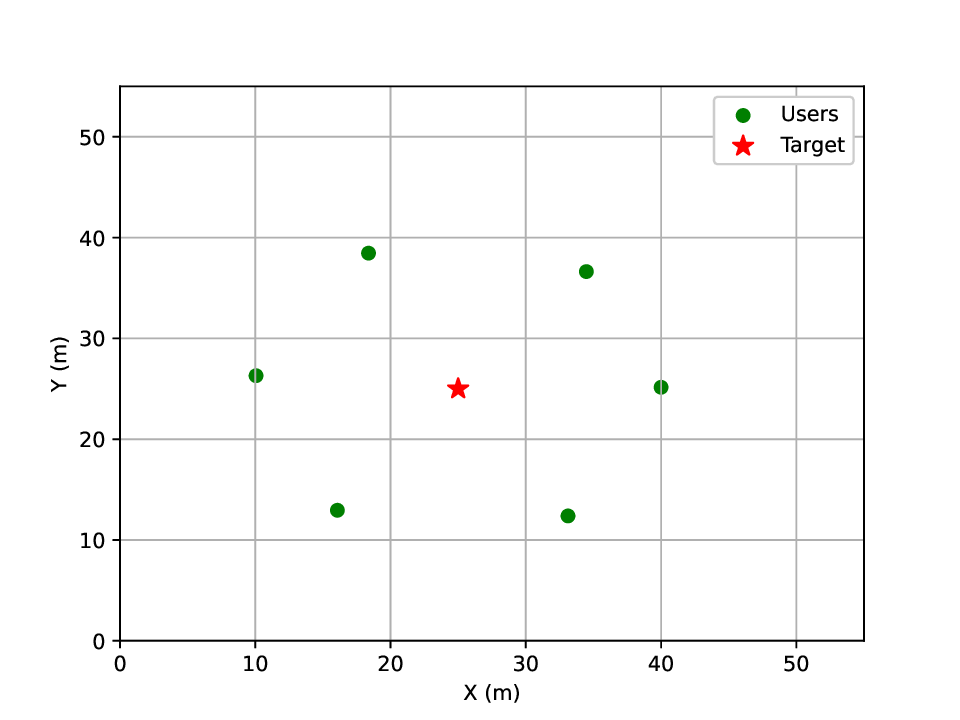}
		\caption{User and Target Positions}
	\end{subfigure}
	\hfill
	\begin{subfigure}[t]{0.24\textwidth}
		\centering
		\includegraphics[width=\linewidth]{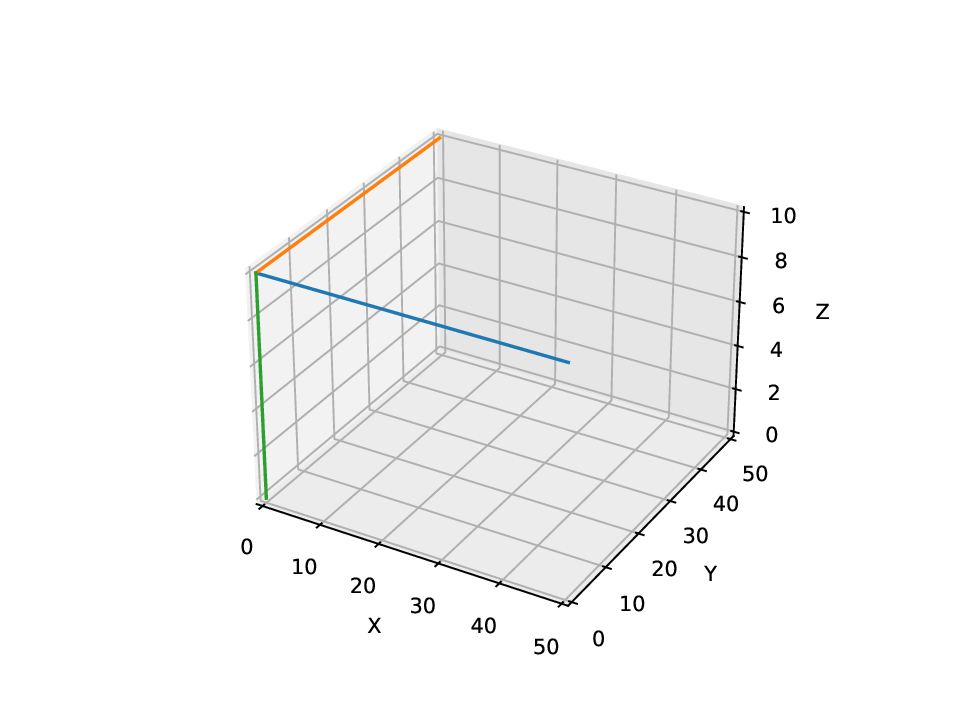}
		\caption{3D Waveguide Deployment}
	\end{subfigure}

	\begin{subfigure}[t]{0.24\textwidth}
		\centering
		\includegraphics[width=\linewidth]{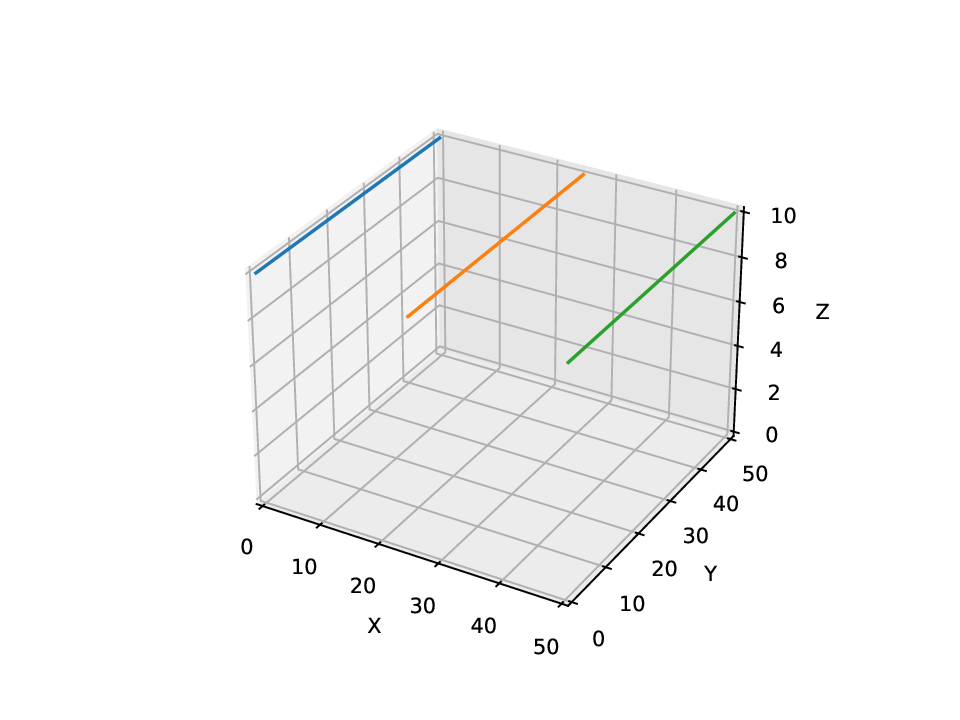}
		\caption{2D Waveguide Deployment}
	\end{subfigure}
	\hfill
	\begin{subfigure}[t]{0.24\textwidth}
		\centering
		\includegraphics[width=\linewidth]{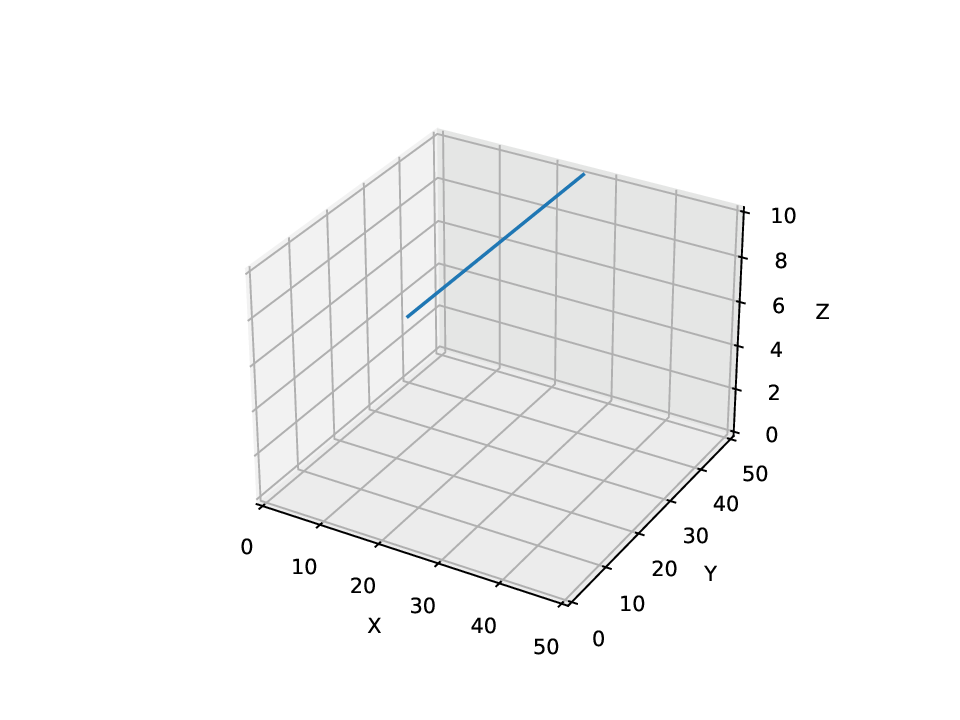}
		\caption{1D Waveguide Deployment}
	\end{subfigure}
	
	\caption{Illustration of simulation configurations.}
	\label{Fig.4}
\end{figure}

To evaluate the performance of different antenna deployment strategies, we construct a simulated 3D environment. A single sensing target is placed near the center of the region, and multiple communication users are randomly positioned along a circular ring surrounding the target, representing a realistic clustered user scenario. Three representative antenna deployment configurations are investigated and illustrated in Fig.~\ref{Fig.4}:

\begin{itemize}
	\item \textbf{1D Deployment:} All antennas are placed along a straight vertical line at a fixed horizontal position ($x = 25\,\text{m}$) and height $z = 10\,\text{m}$, forming a linear array.
	\item \textbf{2D Deployment:} Antennas are arranged in three vertical lines at $x = 0\,\text{m}, 25\,\text{m}, 50\,\text{m}$, all located at a fixed elevation $z = 10\,\text{m}$, simulating a 2D planar distribution.
	\item \textbf{3D Deployment:} Antennas are deployed along three mutually perpendicular lines starting from a common origin at $(0\,\text{m}, 0\,\text{m}, 10\,\text{m})$, extending along the $x$, $y$, and $z$ axes, forming a spatial pinching configuration.
\end{itemize}
This simulation setup enables direct comparison of communication and sensing performance under various spatial layouts, reflecting both practical deployment constraints and theoretical design diversity.

\begin{figure}[htbp]
	\centering
	\captionsetup{justification=centering}
	\includegraphics[width=0.5\textwidth]{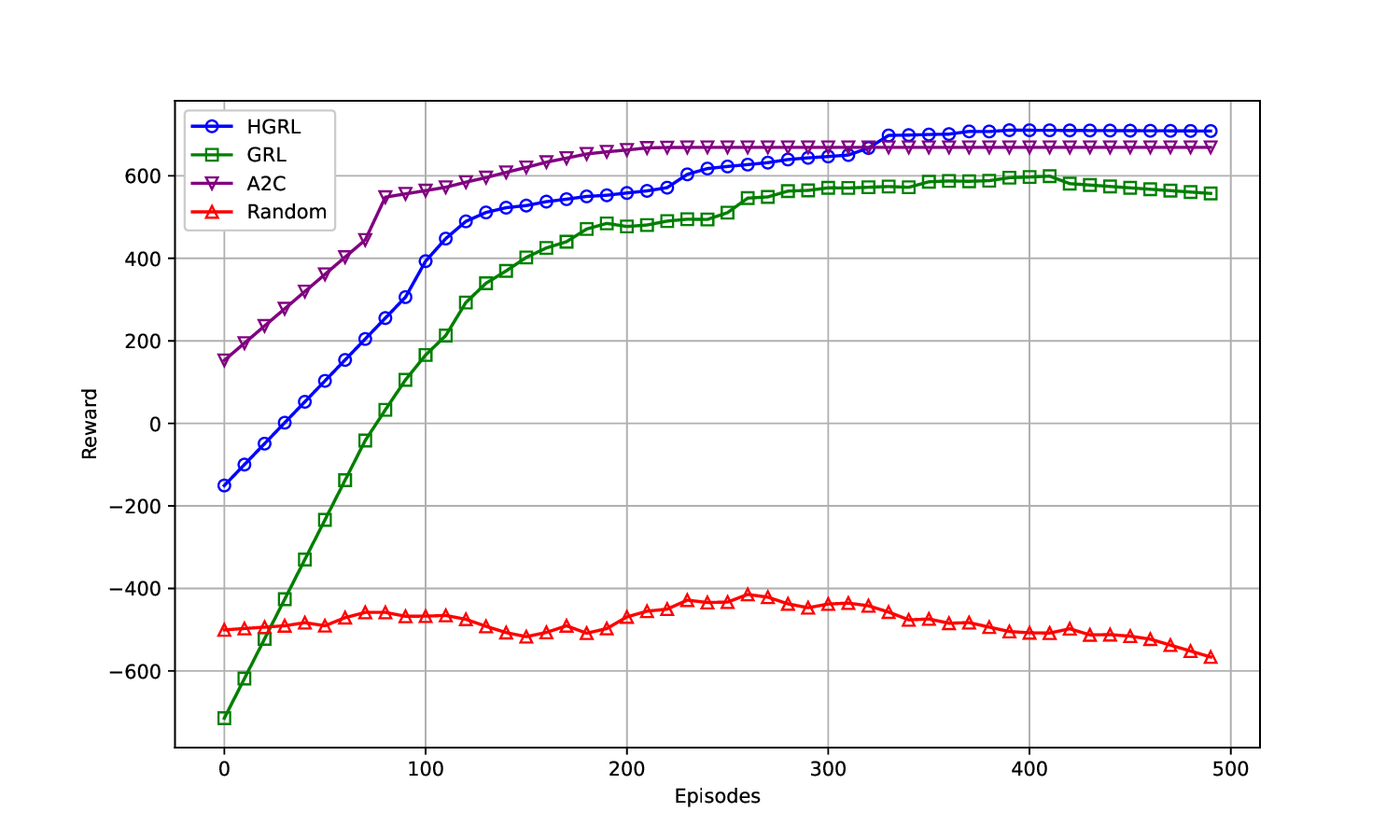}
	\caption{Reward performance of the proposed HGRL compared to baselines.}
	\label{Fig.2}
\end{figure}

Fig. \ref{Fig.2} presents the cumulative reward performance of the proposed HGRL algorithm compared with GRL, A2C, and a baseline of random configuration. Among the baselines, A2C exhibits the fastest convergence, benefiting from its lightweight structure and stable actor-critic updates, but its final performance is limited due to the lack of relational modeling. GRL improves upon A2C by introducing a homogeneous graph representation, enabling better information aggregation across entities, yet it still falls short in overall reward. The proposed HGRL framework, which models the environment using a heterogeneous graph to distinguish between different types of entities and interactions (e.g., communication vs. sensing components), achieves the highest final reward. Although HGRL shows slower convergence due to its more complex hierarchical and graph-based reasoning process, it ultimately outperforms all baselines in terms of long-term performance and reward stability. This demonstrates the effectiveness of heterogeneous relational modeling in capturing the task-specific structure of ISAC systems.

\begin{figure}[t!]
	\centering
	\captionsetup{justification=centering}
	\includegraphics[width=0.5\textwidth]{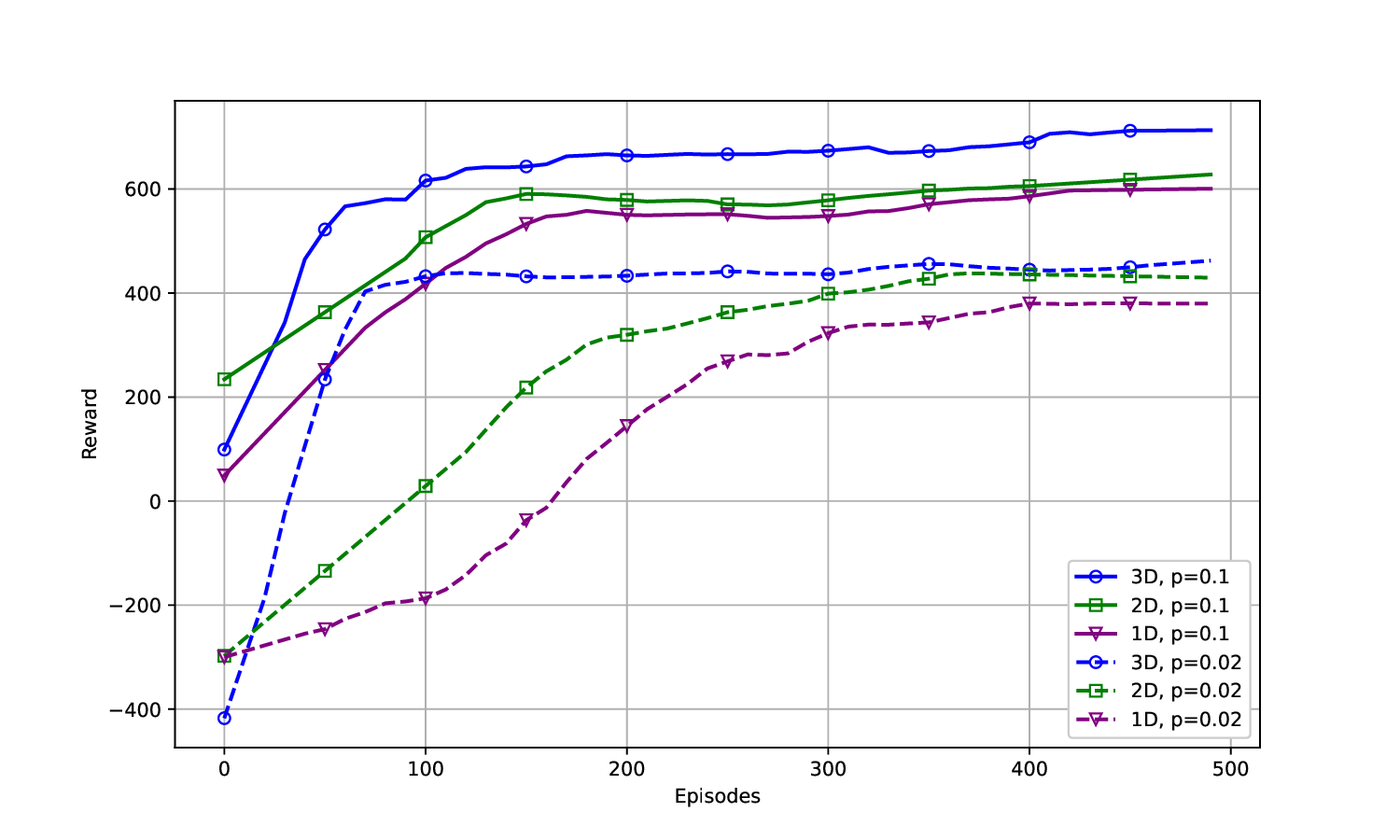}
	\caption{ISAC performance comparison of different pinching antenna deployment.}
	\label{Fig.3}
\end{figure}

Fig.~\ref{Fig.3} illustrates the reward performance under three different pinching antenna deployments with two maximum per-antenna power constraints, $p=0.1\textit{W}$ and $p=0.02 \textit{W}$. In the high-power setting ($p=0.1\textit{W}$), the 3D deployment significantly outperforms the 2D and 1D cases in both convergence speed and final reward, demonstrating the advantage of enhanced spatial diversity. Conversely, when the power is constrained to $p=0.02\textit{W}$, all deployments exhibit slower convergence and lower overall reward. However, the performance gap between the 3D and lower-dimensional deployments becomes more pronounced, emphasizing that 3D deployment is more power-efficient and robust under stricter power limitations.

\begin{table}[ht]
	\centering
	\caption{Comparison of communication and sensing performance under different antenna power and deployment schemes}
	\begin{tabular}{|c|c|c|c|}
		\hline
		\makecell{\textbf{Per antenna} \\ \textbf{power (W)}} & \textbf{deployment} & \makecell{\textbf{Avg. communication} \\ \textbf{rate (bps/Hz)}} & \makecell{\textbf{Avg. sensing} \\ \textbf{SNR (dB)}} \\
		\hline
		\multirow{3}{*}{0.1} & 1D &6.21 &36.49 \\
		& 2D & 6.37 & 19.88 \\
		& 3D & 7.13 & 7.90 \\
		\hline
		\multirow{3}{*}{0.02} & 1D & 3.80 & 17.59 \\
		& 2D & 4.50 & 12.44 \\
		& 3D & 4.70 & 5.41 \\
		\hline
	\end{tabular}
	\label{Table1}
\end{table}

Table~\ref{Table1} compares the average communication rate and sensing SNR achieved under different antenna deployments and maximum power constraints. The results demonstrate that the proposed optimization strategy effectively maximizes the communication rate while ensuring the sensing SNR remains above the required threshold of 5~dB. Notably, the 3D pinching antenna deployment achieves the highest communication performance across both power levels. This is because the 3D structure offers enhanced spatial flexibility, allowing the sensing SNR to closely approach the constraint (e.g., 5.41~dB when $P=0.02$~W), thereby reducing unnecessary power consumption for sensing and allocating more power to communication. In contrast, the 1D and 2D deployments yield significantly higher sensing SNRs, indicating inefficient power usage that limits communication performance. These results highlight the advantage of 3D deployment in striking a better balance between sensing and communication under power-limited conditions.

\section{Conclusion}\label{section:5}
In this paper, we proposed a novel 3D deployment strategy for pinching antenna arrays to enhance ISAC systems. By leveraging the spatial diversity of tri-axial waveguides, the 3D deployment enables flexible beam steering and efficient separation of communication and sensing beams. We formulated the joint optimization of antenna positioning, TDMA allocation, and power assignment as a Markov Decision Process and solved it using a HGRL framework. The proposed HGRL algorithm outperforms baseline methods in both performance and convergence by incorporating a heterogeneous GNN encoder that captures task-specific topology. Simulation results verify that the 3D structure achieves the highest communication rate while maintaining the sensing SNR near the constraint threshold. This ensures optimal use of resources and highlights the superiority of 3D pinching antenna configurations for power-limited ISAC scenarios. 
\bibliographystyle{IEEEtran}
\bibliography{ref}

\end{document}